\documentclass[letterpaper,10pt]{IEEEtran}

\usepackage{amsmath}
\usepackage{amssymb}
\usepackage{epsfig}
\usepackage{balance}
\usepackage{flushend}
\usepackage[noadjust]{cite}

\IEEEoverridecommandlockouts

\author{
\authorblockN{
E. Papagiannis\authorrefmark{1},
C. Tjhai\authorrefmark{1},
M. Ahmed\authorrefmark{2},
M. Ambroze\authorrefmark{1},
M. Tomlinson\authorrefmark{1}\\}
\authorblockA{\authorrefmark{1}Fixed and Mobile Communications Research, University of Plymouth, Plymouth, PL4 8AA, United Kingdom}
\authorblockA{\authorrefmark{2}Centre for Research in Information Storage Technology, University of Plymouth, PL4 8AA, United Kingdom}
email: \{epapagiannis,ctjhai,mahmed,mambroze,mtomlinson\}@plymouth.ac.uk
\thanks{This work was partly funded by the Overseas Research Students award and the Data Storage Network, UK.}
}

\title{Improved Iterative Decoding for Perpendicular Magnetic Recording}

\begin{document}

\maketitle

\begin{abstract}
An algorithm of improving the performance of iterative decoding on perpendicular magnetic recording is presented. This algorithm
follows on the authors' previous works on the parallel and serial concatenated turbo codes and low-density parity-check codes. The
application of this algorithm with signal-to-noise ratio mismatch technique shows promising results in the presence of media noise.
We also show that, compare to the standard iterative decoding algorithm, an improvement of within one order of magnitude can be achieved. 
\end{abstract}

\section{Introduction}\label{sec:introduction}
Longitudinal recording has been the standard method in magnetic recording for decades. Recent research has shown that this method seems
to reach its physical limits in the near future due to the superparamagnetic effects. On the other hand, the technique which has been
known prior to the longitudinal recording--perpendicular recording, has recently been the centre of research attention. Perpendicular
magnetic recording offers promising increased in recording densities, up to 1 Terabit per square inch seems feasible~\cite{Wood.2000}.
As the areal density is increased, however, the signal processing aspects of magnetic recording becomes more difficult. Sources of
distortion including media noise, electronics and head noise, jitter noise, inter-track interference, thermal asperity, partial erasure
and dropouts become more apparent and unless appropriate mitigation techniques are present, signals cannot be retrieved reliably from the
recording media.

Since the discovery of turbo codes, soft-decision iterative decoding has been shown to be able to provide significant coding gain over the
conventional detection method on magnetic recording. The utilisation of iterative decoding on the concatenation of partial-response (PR)
channel and powerful error-correcting codes such as low-density parity-check (LDPC) and turbo codes has been proposed in many literatures.
Iterative decoding is a reduced-complexity method to achieve the optimum solution--the maximum-likelihood solution and as such, iterative
decoding is sub-optimal.

In this paper, we present a method to improve the sub-optimality of the iterative decoding and demonstrate its applications to perpendicular
magnetic recording in the presence of media noise. The improved method, which is known as the Received-Vector-Coordinate-Modification (RVCM)
algorithm, follows on the previous works of the authors~\cite{Papagiannis_et_al.patent},\cite{Papagiannis_et_al.2003},\cite{Papagiannis_et_al.2004},
\cite{Tjhai_et_al.patent},\cite{Papagiannis_et_al.isit2005}. This method is similar to the works of~\cite{Pishro_et_al.2003} and~\cite{Varnica_et_al.2004}.
This paper also investigates the use of signal-to-noise ratio (SNR) mismatch~\cite{Tan_et_al.2004} to mitigate the effect of  media
noise.

The rest of the paper is organised as follows. Section~\ref{sec:channel} describes the perpendicular recording channel used. The description
of the RVCM algorithm is outlined in Section~\ref{sec:rvcm} and the performance of this algorithm is demonstrated in Section~\ref{sec:performance}.
Section~\ref{sec:conclusions} concludes this paper.

\begin{figure*}[th!]
\centering\includegraphics[width=5in]{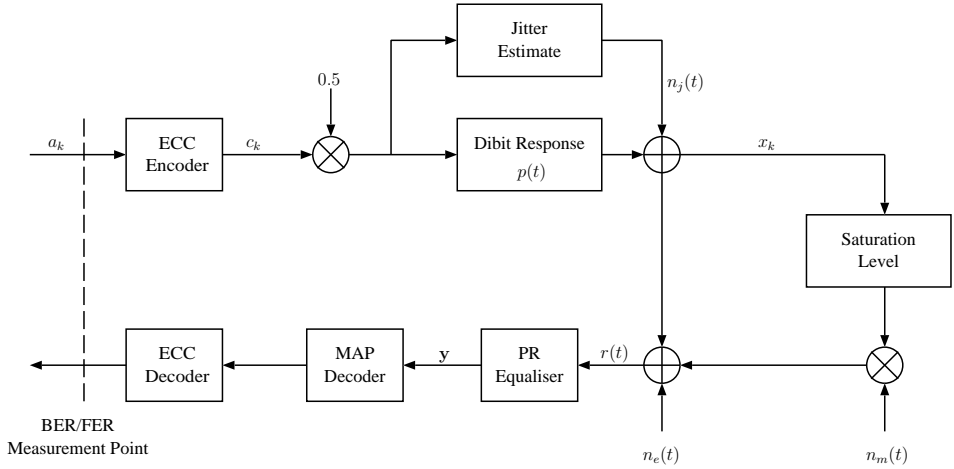}
\caption{\label{fig:block-diagram}Perpendicular recording system model}
\end{figure*}

\section{Perpendicular Recording Channel Model}\label{sec:channel}
Fig.~\ref{fig:block-diagram} shows the block diagram of the perpendicular recording system model used in this paper.
The user data, denoted as $a_k$, is a sequence of of input symbols taking values of $\{0,1\}$. Some error-protection
redundancy is added to the sequence $a_k$ by the error-correcting-codes (ECC) encoder forming codeword sequence $c_k$.
To simulate the write current, the sequence $c_k$ is mapped to $\{-1,+1\}$ according to $2c_k-1$ operation. The scaling
factor of $0.5$ is to ensure the transition takes values of $\{-1,0,+1\}$.

We assume that the read head produces zero voltage in the region of magnetic transitions and some voltage in the region of constant magnetic polarity.
We approximate the single-transition step response, denoted as $s(t)$, using the hyperbolic-tangent function~\cite{Sawaguchi_et_al.2001}:
\begin{align}
s(t) &= A\cdot\tanh\left(\ln(3)\frac{t}{\text{PW}_{50}}\right)\label{eqn:st}
\end{align}
where $A$ is the saturation level or the amplitude from zero to peak (normalised to unity) and $\text{PW}_{50}$ is the time taken for $s(t)$ to go
from $-A/2$ to $+A/2$. It is assumed that $t$ and $\text{PW}_{50}$ are normalised to the symbol period, $T$. 
Throughout the paper, it is assumed that $\text{PW}_{50} = 1.4$. We define the response of two adjacent transitions (dibit-response) $p(t)$ as:
\begin{align}
p(t) &= s(t) - s(t-1)\label{eqn:pt}
\end{align}
and the readback signal $r(t)$ is simply the convolution of $c_k$ and $p(t)$ plus some noise:
\begin{align}
r(t) &= \sum_k c_k p(t-kT) + n(t)\label{eqn:rt}
\end{align}
where $n(t)$ is the overall noise in the recording system which consists of media, jitter and electronic noise, i.e. $n(t) = n_m(t) + n_j(t) + n_e(t)$.

The media noise, $n_m(t)$ originates from the imperfections of the media and its effect is significant in the magnetic transition regions.
Typically, media noise is approximately four times the electronic noise at transition regions. In our system model, we consider the media
noise as Additive-White-Gaussian-Noise (AWGN) with mean of $0$ and variance of $\sigma^2_m$, which exists in the transition region only.
As shown in Fig.~\ref{fig:block-diagram}, the media saturation noise depends on the saturation level and it is evaluated as $1-(x_k/A)^2$. 
Unlike media noise which is media dependent, the jitter noise $n_j(t)$ is due to timing imperfection only. To model the sampling jitter noise,
$n_j(t)$, the $n$th order Taylor approximation is used. The jitter estimation block shown in Fig.~\ref{fig:block-diagram} is done with $6$th
order Taylor series expansion of $s(t)$. The jitter probability density function is assumed to be uniform, limited by a maximum value.
The electronic noise, $n_e(t)$ is AWGN with mean of $0$ and variance of $\sigma_e$. The recording system in Fig.~\ref{fig:block-diagram}
caters for many different simulation cases with varying degree of electronics, media and sampling jitter noise. We define the channel SNR as:
\begin{align}
\text{SNR} &= 10\log_{10}\left(\frac{1}{2\left(\sigma_e^2 + \sigma_m^2\right)}\right)\label{eqn:snr}
\end{align}

The noisy readback signal is equalised to $(4,6,4,2)$ PR target which is only optimal for electronic noise at the considered
$\text{PW}_{50}$~\cite{Madden_et_al.2004}. It serves for comparison purposes only. The Maximum-a-Posteriori (MAP) decoder of the PR channel
exchanges extrinsic with the ECC decoder to deliver solution which is used for performance evaluation.

\section{Received-Vector-Coordinate-Modification (RVCM) Algorithm}\label{sec:rvcm}
It has been shown that the RVCM algorithm provides considerable coding gain for parallel and serial concatenated turbo
codes~\cite{Papagiannis_et_al.2003},\cite{Papagiannis_et_al.2004} and LDPC codes~\cite{Papagiannis_et_al.isit2005}. The algorithm can be
applied directly to perpendicular recording and is described briefly below.

\subsection{Description of the Algorithm}
Let $\mathbf{y}=\{y_0,y_1,\ldots,y_n\}$ denote an $n-$tuple vector at the output of the MAP decoder, that is the a-posteriori probability (APP)
of the MAP decoder. Let $\mathbf{l}=\{l_0,l_1,\ldots,l_n\}$ denote the reliability sequence of $\mathbf{y}$, where
$l_i=\log\left(\text{Pr}(y_i|+1)/\text{Pr}(y_i|-1)\right)$. Assume that $i_{\text{max}}$ is an integer where $1 \le i_{\text{max}} \le n$ and
$\mathbf{p} \subseteq \{0,1,\ldots,n-1\}$ is a vector of length $i_{\text{max}}$.
\begin{description}[\setlabelwidth{Step XX.}]
\item [\textbf{Step 1.}]  Store the vector $\mathbf{l}$, let the integer $i$ be initialised to $0$.
\item [\textbf{Step 2a.}] Set $l^\prime_{\mathbf{p}_i} = l_{\mathbf{p}_i}$ and $l_{\mathbf{p}_i} = -\infty$. 
Restart the iterative decoder, store the decoded vector ($\mathbf{d^-_{\mathbf{p}_i}}$).
\item [\textbf{Step 2b.}] Set $l_{\mathbf{p}_i} = +\infty$. 
Restart the iterative decoder, store the decoded vector ($\mathbf{d^+_{\mathbf{p}_i}}$) and
restore $l_{\mathbf{p}_i}$, i.e. $l_{\mathbf{p}_i} = l^\prime_{\mathbf{p}_i}$.
\item [\textbf{Step 3.}]  If $i < i_{\text{max}}$ then set $i=i+1$ and continue to \textbf{Step 2}. Otherwise, stop the
algorithm and from the list of all decoded vectors $\mathbf{d^-_{\mathbf{p}_i}} \bigcup \mathbf{d^+_{\mathbf{p}_i}}$,
$\forall i \in \{0,1,\ldots,i_{\text{max}}-1\}$, choose a decoded vector that has the minimum euclidean distance.
It is assumed that the iterative decoder always outputs a codeword.
\end{description}
From the steps above, it is clear that the complexity of the algorithm depends on $i_{\text{max}}$.

\subsection{Critical Symbols}
One of the major obstacles concerning the RVCM algorithm is the difficulty in finding the symbol(s) that, if modified, can converge the
iterative decoder to the maximum-likelihood solution~\cite{Papagiannis_et_al.2003},\cite{Papagiannis_et_al.2004}. These symbols are referred
as the critical symbols and their distribution is uniform with no sign of vulnerable or favourite symbol positions. On
the other hand, due to their uniform distribution, it is likely that we can find one of the critical symbols if we confine our
search to a small group, i.e. keeping the value of $i_{\text{max}}$ low. In this way, we can reduce the computational complexity for the price
of sub-optimum performance. As we will show later that, the gain obtained by confining $i_{\text{max}}$ to a small value is still significant
compared to the performance of the standard iterative decoder.

There are various methods for selecting the critical symbols, see~\cite{Papagiannis_et_al.isit2005} for details. In this paper, we restrict
the selection to one method only, that is the reliability of the APP at the output of the MAP decoder.

\section{RVCM Performance}\label{sec:performance}
\begin{figure}[!t]
\centering\includegraphics[angle=270,width=2.3in]{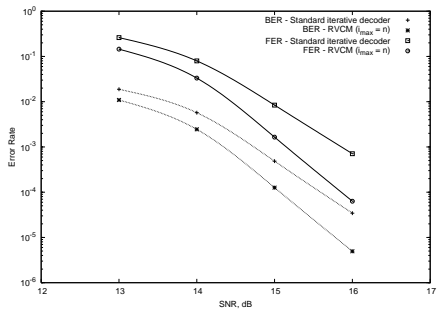}
\caption{\label{fig:turbo-Sm0}Error performance of RVCM decoder on the turbo code in the presence of electronic noise only}
\end{figure}

\begin{figure}[!t]
\centering\includegraphics[angle=270,width=2.3in]{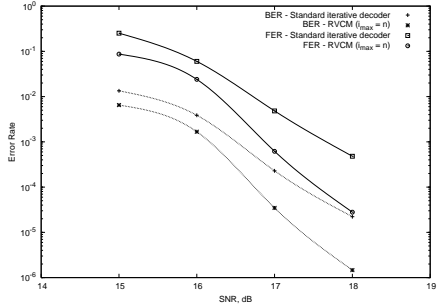}
\caption{\label{fig:turbo-Sm4}Error performance of RVCM decoder on the turbo code in the presence of electronic and media noise}
\end{figure}

We evaluate the performance of the RVCM algorithm on some short-block length turbo and LDPC codes. Fig.~\ref{fig:turbo-Sm0}
and~\ref{fig:turbo-Sm4} show the error rate performance of the turbo code under the standard iterative and RVCM decoders.
The turbo code considered is the $b/b+1$ tail-biting turbo code\footnote{C. Berrou once referred this code as duo-binary turbo code.}, 
where $b=3$, $k=198$ and $n=278$ with $\mathcal{S}-$interleaver.
Significant improvement is noticed and the increase in performance gets better as SNR increases. It is worth noting that the results
in Fig.~\ref{fig:turbo-Sm4} were obtained using SNR mismatch technique. With this technique, the improvement in performance
over standard iterative decoder is even greater as SNR increases.
In the presence of media noise, SNR mismatch methods do not provide the same performance as observed with electronics noise
only~\cite{Tan_et_al.2004}, however better targets for media noise are being investigated by the authors.

\begin{figure}
\centering\includegraphics[angle=270,width=2.3in]{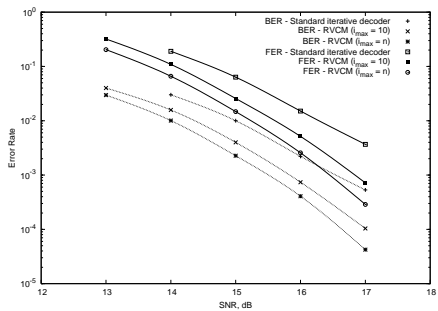}
\caption{\label{fig:gf2-127-84}Error performance of RVCM decoder on the $(127,84)$ cyclic LDPC code in the presence of electronic and media noise}
\end{figure}
\begin{figure}
\centering\includegraphics[angle=270,width=2.3in]{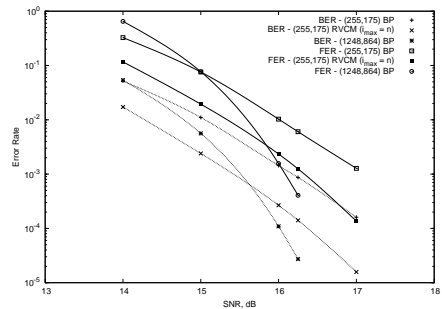}
\caption{\label{fig:gf2-255-175}Performance of the $(255,175)$ cyclic and $(1248,864)$ codes in the presence of electronic and media noise}
\end{figure}

Similar performance improvement is observed for the LDPC codes, see Fig.~\ref{fig:gf2-127-84}. The $(127,84)$ cyclic LDPC code,
which has minimum-distance of $9$, was constructed using a method described in~\cite{Horan_et_al.itw2005}. As mentioned earlier,
the RVCM algorithm allows one to trade off the performance against the computational complexity. From Fig.~\ref{fig:gf2-127-84},
despite the performance obtained by setting $i_{\text{max}}=10$ is approximately $0.3$dB inferior to that by setting
$i_{\text{max}}=n$, the coding gain from the standard iterative decoding is significant. For the case of $i_{\text{max}}=10$, we
select the critical symbols based on the reliability measure at the output of the MAP decoder. 
From the vector $\mathbf{l}$, we construct a vector $\mathbf{p}$ of length $i_{\text{max}}$ such that
$l_{\mathbf{p_0}} < l_{\mathbf{p}_1} < l_{\mathbf{p}_2} < \ldots < l_{\textbf{p}_{n-1}}$.
In Fig.~\ref{fig:gf2-255-175}, we compare the performance of the $(255,175)$ cyclic code and that of the $(1284,864)$ quasi-cyclic code.
We can see that the RVCM algorithm provides significant gain, within one order of magnitude improvement, over the BP algorithm. At
approximately $3\times10^{-5}$ BER, the performance of the cyclic code with RVCM is within $0.6$dB away from the longer code under BP decoding.

\section{Conclusions}\label{sec:conclusions}
We have shown that the application of the RVCM algorithm to perpendicular magnetic recording shows promising results. Simulation results
show that improvement of within one order of magnitude is possible. Short block length offers an attractive error-correction scheme in
which RVCM algorithm can be fully exploited by setting $i_{\text{max}}=n$. 
A bank consisting of $2i_{\text{max}}$ parallel RVCM decoders can be built on chips and the decoding
of short-block length data has low latency. 
The performance of longer block-length codes, up to a certain error-rate, can be outperformed by the application of RVCM algorithm to
shorter codes. The exact point, at which the longer codes start to perform better, depends on the code structure.

We also extended our investigations on using some non binary cyclic LDPC codes~\cite{Tjhai_et_al.isit2005} and we observe similar improvement
as in the binary cases. Further investigations in identifying the critical symbols will allow the application of RVCM algorithm to long powerful
codes.

\section*{Acknowledgement}
\addcontentsline{toc}{section}{Acknowledgement}
The authors would like to thank Prof. Barry K. Middleton of University of Manchester for the channel noise discussion.

\end{document}